\documentclass[prl,twocolumn,showpacs,amsmath,amssymb]{revtex4}
\usepackage{graphicx}% Include figure files
\usepackage{dcolumn}% Align table columns on decimal point
\usepackage{bm}% bold math

\newcommand{\psip}{\psi(2S)}

\newcommand{\jpsi}{J/\psi}

\newcommand{\EE}{e^+e^-}
\newcommand{\MM}{\mu^+\mu^-}

\newcommand{\OP}{\omega\pi^0}
\newcommand{\KKSC}{K^{*+}K^-}
\newcommand{\KKSN}{K^{*0}\overline{K^0}}

\newcommand{\ra}{\rightarrow}

\newcommand{\rhopi}{\rho\pi}
\newcommand{\beq}{\begin{equation}}
\newcommand{\eeq}{\end{equation}}
\newcommand{\bfg}{\begin{figure}}
\newcommand{\efg}{\end{figure}}
\newcommand{\bitm}{\begin{itemize}}
\newcommand{\eitm}{\end{itemize}}
\newcommand{\bnum}{\begin{enumerate}}
\newcommand{\enum}{\end{enumerate}}
\newcommand{\btbl}{\begin{table}}
\newcommand{\etbl}{\end{table}}
\newcommand{\btbu}{\begin{tabular}}
\newcommand{\etbu}{\end{tabular}}
\newcommand{\aggg}{a_{3g}}
\newcommand{\agcc}{a_{\gamma}}
\newcommand{\agee}{a_c}
\newcommand{\VP}{1^-0^-}

\begin{document}

\title{Possible Large Phase in $\psi(2S) \rightarrow 1^-0^-$ Decays}

\author{P.Wang$^{1}$}
\email{wangp@mail.ihep.ac.cn}
\author{C.Z.Yuan$^{1}$}
\author{X.H.Mo$^{1,2}$}
 
\affiliation{
$^1$Institute of High Energy Physics, CAS, Beijing 100039, China\\
$^2$China Center of Advanced Science and Technology (World Laboratory),
Beijing 100080, China}
 
\date{\today}
	   
\begin{abstract}
The strong and the electromagnetic amplitudes are analyzed on the basis of
the measurements of $\jpsi, \psi(2S)\rightarrow \VP$ in $\EE$ experiments.
The currently available experimental information is revised with inclusion
of the contribution from $\EE \rightarrow \gamma^* \rightarrow \VP$. 
The study shows that a large phase around $-90^{\circ}$ between the strong
and the electromagnetic amplitudes could not be ruled out by the experimental 
data for $\psip$.
\end{abstract}
\pacs{12.38.Aw, 13.25.Gv, 13.40.Gp, 14.40.Gx}% PACS
\maketitle

In recent years, the increased information on $\jpsi$ and $\psip$ decays 
from experiments has led to the analysis of strong and electromagnetic
decay amplitudes in charmonium decay
processes~\cite{suzuki,dm2exp,mk3exp,a00,a11,ann}. Such analysis on
$\jpsi$ revealed that there exists a relative orthogonal phase between these
two amplitudes for the two-body decay modes: $1^+0^-$~\cite{suzuki},
$1^-0^-$~\cite{dm2exp,mk3exp}, $0^-0^-$~\cite{a00,a11}, $1^-1^-$~\cite{a11}
and $N\overline{N}$~\cite{ann}. 

As to the $\psip$ decay for which the information from experiments is less
abundant than $\jpsi$,  it is a question whether it decays in the same
pattern. It has been argued~\cite{suzuki} that the only large energy scale
involved in the three-gluon decay of charmonia is the charm quark mass, one
expects that the corresponding phase should not be much different between
$\jpsi$ and $\psip$ decays. There is also another theoretical argument which
favors the $\pm90^\circ$ phase~\cite{gerard}. This large phase follows from  
the orthogonality of three-gluon and one-photon virtual processes. But an
extensively quoted work~\cite{suzuki} found that a fit to $\psip\rightarrow
1^-0^-$ with a large phase $\pm90^\circ$ is virtually impossible and
concluded that the relative phase between strong and electromagnetic
amplitudes should be around $180^\circ$~\cite{refft1}.
So it is a matter of great concern that whether the large phase is
consistent with the $\psip$ experimental data. 

Up to now, the most accurate data on $\psip \rightarrow 1^-0^-$ are from
$\EE$ colliding experiments. However, in previous analyses, the
contribution from the continuum one-photon annihilation 
$$\EE \rightarrow \gamma^* \rightarrow 1^-0^- $$
has been neglected~\cite{wymzprl}. In this analysis, such contribution will be
taken into account for both $\jpsi$ and $\psip$. First the available data 
from $\EE \rightarrow \jpsi$ are re-analyzed. To avoid the complexity and
uncertainty of the mixing between SU(3) singlet and octet, only four
processes are used, to wit 
\beq
\begin{array}{ccl}
\EE &\rightarrow& \OP~~, \\
\EE &\rightarrow& \rhopi~~, \\
\EE &\rightarrow& \KKSC + c.c.~~,\\
\EE &\rightarrow& \KKSN + c.c.~~.
\end{array}
\eeq
It is found that the phase between strong and electromagnetic decay
amplitudes is either  $-72.0^{\circ}$ or $+76.8^{\circ}$. Then with the same
scheme, the data on $\EE \rightarrow \psip$ are re-examined. 
It is found that the currently available data from BES~\cite{bes},
accommodate the phases of both $180^{\circ}$ and $-90^{\circ}$. 

In $\EE\rightarrow1^-0^-$ at $\jpsi$ or $\psip$ resonance, the Born order
cross section for final state $f$ is
\beq
\sigma_{Born} = 
\frac{4\pi\alpha^2}{s^{3/2}}|A_f|^2{\cal P}_f(s)~~,
\label{Born}
\eeq
where ${\cal P}_{f}(s) = q_f^3/3$, with $q_f$ being the momentum of either the
$1^-$ or the $0^-$ final state particle. 

In $\EE$ annihilation experiment, there are three
amplitudes~\cite{wymzprl,rudaz}: the continuum one-photon annihilation
amplitude $a_c$, the electromagnetic decay amplitude of the resonance
$\agcc$ and the strong decay amplitude of the resonance $\aggg$. For the
SU(3) breaking processes, a SU(3) breaking term $\epsilon$ is added to
$\aggg$, so the strong decay amplitude is $\aggg+\epsilon$. With inclusion
of $a_c$, the amplitudes of the four $\EE\rightarrow1^-0^-$ processes are
expressed as: 
\beq
\begin{array}{ccl}
A_{\OP} &=& 3(a_\gamma+a_c)~~ , \\
A_{\rhopi} &=& \aggg+a_\gamma+a_c~~ , \\
A_{\KKSC} &=& \aggg+\epsilon+a_\gamma+a_c~~ , \\
A_{\KKSN} &=& \aggg+\epsilon-2(a_\gamma+a_c) ~~.
\end{array}
\label{arp}
\eeq
For $\OP$ which goes only through electromagnetic process, $a_c$ and 
$a_\gamma$ are related to the $\OP$ form factor ${\cal F}_{\OP}(s)$: 
\beq
a_c =  \frac{1}{3}{\cal F}_{\OP}(s)~~,
\label{ac}
\eeq
and
\beq
a_\gamma = \frac{\sqrt{s}\Gamma_{ee}/\alpha}{s-M^2+iM\Gamma_t}
{\cal F}_{\OP}(s)~~,
\label{ag}
\eeq
where $\alpha$ is the QED fine structure constant, $M$ and $\Gamma_t$ are
the mass and total width of $\jpsi$ or $\psip$, and $\Gamma_{ee}$ is the
partial width of $\EE$. It has been assumed that there is no extra phase between
$\agee$ and $\agcc$, as in $\EE \rightarrow \MM$.
Formally, the amplitude of $\OP$ process could be written as 
\beq
A_{\OP} = (1 + B(s)){\cal F}_{\OP}(s)~~,
\label{aop}
\eeq
with the definition
$$B(s) \equiv \frac{3\sqrt{s}\Gamma_{ee}/\alpha}{s-M^2+iM\Gamma_t}~~.$$
If there is no $a_c$ in Eq.(\ref{aop}), only the second term is left which
describes the resonance decaying through electromagnetic process. 
Substituting it into Eq.~(\ref{Born}), the commonly known Breit-Wigner form
is then reproduced 
$$
\sigma_{BW}(\EE\rightarrow \mbox{Res.} \rightarrow\OP)\;
=\frac{12\pi \Gamma_{ee} \Gamma_{\OP}}{(s-M^{2})^{2}
+\Gamma^{2}_t M^{2}}~~, $$
with 
$$\Gamma_{\OP} =  \frac{\Gamma_{ee}
q^3_{\OP}}{M} |{\cal F}_{\OP}(M^2)|^2~~.$$

For the strong decay amplitude, the most interesting point lies in its 
phase and strength relative to the electromagnetic decay amplitude, 
so it is parametrized in the way:
\beq
\aggg = {\cal C}e^{i\phi}a_\gamma~~, 
\label{aggg}
\eeq
where $\phi$ is the phase between the two amplitudes and ${\cal C}$
is taken to be real. For the SU(3) breaking strong decay amplitude, it
is parametrized as its strength relative to the SU(3) conserved one:
\beq
{\cal R}= \frac{\aggg+\epsilon}{\aggg}~~.
\label{aggge}
\eeq
As in Refs.~\cite{suzuki,haber,bramon}, it is assumed that the SU(3) breaking 
amplitude $a_{3g}+\epsilon$ has the same phase as $\aggg$~\cite{refft2},
so ${\cal R}$ is real. According to Eqs.~(\ref{aggg}) and (\ref{aggge}),
together with Eq.~(\ref{aop}), the amplitudes of Eq.~(\ref{arp}) could be 
expressed as:
\beq
\begin{array}{ccl}
A_{\OP}   &=& [1 + B(s)] \cdot {\cal F}_{\OP}(s) ~~, \\
A_{\rhopi}&=& [({\cal C}e^{i\phi}+1)B(s)+1 ] \cdot {\cal F}_{\OP}(s)/3~~, \\
A_{\KKSC} &=& [({\cal C} {\cal R} e^{i\phi}+1)B(s)+1 ] 
                                             \cdot {\cal F}_{\OP}(s)/3~~, \\ 
A_{\KKSN} &=& [({\cal C} {\cal R} e^{i\phi}-2)B(s)-2 ] 
                                             \cdot {\cal F}_{\OP}(s)/3~~. \\
\end{array} 
\label{arpnew}
\eeq

In this analysis, the branching ratios are converted into measured cross
sections by multiplying the total resonance cross section. Special attention
should be paid in calculating the cross sections where the experimental
conditions must be taken into account properly~\cite{wymhepnp,wymplb}.
The most important ones are the radiative correction and the energy spread
of the collider, both of which reduce the height of the resonance and shift
the position of the maximum cross section. Also experiments naturally tend to 
collect resonance data at the energy which yields the maximum inclusive
hadron cross sections. This energy is higher than the nominal resonance
mass, and it does not necessarily coincide with the maximum cross section of
each exclusive mode. All these must be considered accordingly. 

The experimental results for $\jpsi$ decays relevant to the forementioned four
channels are listed in Table~\ref{multiexpres}. The values of energy spread
are obtained from Ref.~\cite{a11}. The positions which yield the maximum
inclusive hadronic cross section on each $\EE$ collider are calculated and
listed in Table~\ref{multiexpres} as well. 

\begin{table*}
\caption{\label{multiexpres} Experimental results 
for $\EE \rightarrow 1^-0^-$ processes at $\jpsi$ energy region.}
\begin{ruledtabular}
\begin{tabular}{llcccc}
Experiment& Accelerator & C.M. Energy & Data Taking & final state 
          &Branching Ratio                           \\              
          &             &Spread (MeV) &Position\footnote{
\begin{minipage}{13cm}\mbox{}The data taking position is the energy which
yield the maximum inclusive hadronic cross section. \end{minipage} } (GeV)
                                                    & 
          &                                          \\ \hline  
%%%%%%%%%%%%%%%%%%%%%%%%%%%%%%%%%%%%%%%%%%%%%%%%%%%%%%%%%%%%%%%%%%%%%%%%%%%%
DMII~\cite{dm2exp}
          &     DCI     &    1.98     & 3.09707     & $\OP$
          & $(0.0272\pm0.0021)\cdot\cal{B}_{\rhopi}$ \\ 
          &             &             &             & $\KKSC + c.c.$
          & $(0.364\pm0.013)\cdot\cal{B}_{\rhopi}$ \\
          &             &             &             & $\KKSN + c.c.$
          & $(0.300\pm0.011)\cdot\cal{B}_{\rhopi}$ \\
MARK III~\cite{mk3exp} 
          &   SPEAR     &   2.40      & 3.09711     & $\OP$
          & $(4.82\pm 0.52\pm0.41)\times 10^{-4}$   \\
          &             &             &             & $\KKSC + c.c.$
          & $(5.26\pm 0.32\pm0.45)\times 10^{-3}$   \\
          &             &             &             & $\KKSN + c.c.$
          & $(4.33\pm 0.29\pm0.37)\times 10^{-3}$   \\
          &             &             &             & $\rhopi$
          & $(1.42\pm 0.15\pm0.12)\%$              \\
MARK II~\cite{mk2exp}  
          &   SPEAR     &             &             & $\rhopi$
          & $(1.3\pm 0.3)\%$              \\
MARK I~\cite{mk1exp}   
          &   SPEAR     &             &             & $\rhopi$
          & $(1.3\pm 0.3)\%$              \\
CNTR~\cite{cntrexp}     
          &  DORIS      &  1.41       & 3.09701     & $\rhopi$
          & $(1.0\pm 0.2)\%$              \\
PLUTO~\cite{plutoexp}    
          &  DORIS      &             &             & $\rhopi$
          & $(1.6\pm 0.4)\%$              \\
DASP~\cite{daspexp}     
          &  DORIS      &             &             & $\rhopi$
          & $(1.16\pm 0.16)\%$\footnote{\begin{minipage}{17.7cm}
The latest PDG value of ${\cal B}_{\mu\mu}$~\cite{PDG} is used to
renormalize the branching ratio 
${\cal{B}}_{\rhopi}=(1.36\pm 0.28)\% \cdot
\frac{{\cal{B}}_{\mu\mu}=(5.88\pm0.10)\%(\mbox{PDG2002})} 
     {{\cal{B}}_{\mu\mu}=(6.9\pm0.9)\%(\mbox{used by DASP})}.$ 
\end{minipage}}  \\
BES~\cite{besexp}       
          &  BEPC       &   0.85      & 3.09696     & $\rhopi$
          & $(1.21\pm 0.20)\%$              \\ 
\end{tabular}
\end{ruledtabular}
\end{table*}

Chi-square method is employed to fit the experiment data. The 
estimator is defined as
\begin{widetext}
\begin{equation}
\chi^2=\sum\limits_i
\frac{[R_i - \widehat{R}_i(\vec{\eta})]^2}{\sigma^2_i}
      +\sum\limits_j
\frac{[{\cal{B}}_j/f_{mk3} - \widehat{\cal B}_j(\vec{\eta})]^2}
     {(\sigma_j/f_{mk3})^2}
      +\sum\limits_k
\frac{[{\cal{B}}_k - \widehat{\cal B}_k(\vec{\eta})]^2}{\sigma_k^2}
      +\sum\limits_m
\frac{[{\cal{B}}_m - \widehat{\cal B}_m(\vec{\eta})]^2}{\sigma_m^2}
      +\sum\limits_n
\frac{[{\cal{B}}_n - \widehat{\cal B}_n(\vec{\eta})]^2}{\sigma_n^2}~~.
\label{chisqvp}
\end{equation}
\end{widetext}
In above equation, summation index $i$ indicates the results 
from DM II~\cite{dm2exp}; $j$ from MARK III~\cite{mk3exp};  $k$ from MARK
II~\cite{mk2exp} and MARK I~\cite{mk1exp}; $m$ from CNTR~\cite{cntrexp},
PLUTO~\cite{plutoexp} and DASP~\cite{daspexp}; and $n$ from
BES~\cite{besexp}. $R_i$ indicates relative branching ratio, i.e. 
$ R_i = {\cal B}_i / {\cal B}_{\rhopi}$, where $i$ denotes $\OP$,
$\KKSC+c.c.$ and $\KKSN+c.c.$~. The capped symbol in Eq.~(\ref{chisqvp})
indicates the theoretical expectation; and $\vec{\eta}$ denotes the
parameter vector with five elements, four of which have been described in
Eq.~(\ref{arpnew}), and the fifth $f_{mk3}$ is introduced to 
describe the correlation of data from MARK III, and correspondingly the
8.5\% common error (the second term of error) for MARK III measurements in
Table~\ref{multiexpres}, is subtracted from the systematic uncertainty
$\sigma_j$ in Eq.~(\ref{chisqvp}). 

The fitting gives a $\chi^2$ of 4.1 with the number of degrees of freedom
being 7. There are two minima with $\phi$ of opposite sign, while all other
parameters have the same values up to the significant digits listed below: 
\[
\begin{array}{rcl}
   \phi  &=&-72.0^\circ\pm3.6^\circ~, \mbox{or} ~+76.8^\circ\pm3.6^\circ~~; \\
{\cal C} &=& 10.3 \pm0.3~~; \\   
{\cal R}  &=& 0.775 \pm 0.013~~;\\
|{\cal F}_{\OP}(M^2_{\jpsi})|
         &=& (0.075\pm0.004) \mbox{GeV}^{-1}~~. 
\end{array}
\]
$f_{mk3}$ is $1.26\pm 0.11$ from the fit, which means a global
deviation of the MARK III values from other experiments. Above fitting
results of $\jpsi \rightarrow 1^-0^-$ decay deviate little from previous
analysis without $\agee$~\cite{dm2exp,mk3exp}, but the precision is
improved. They support the following theoretical
postulates~\cite{suzuki,gerard}: 
\begin{enumerate}
 \item The relative phase between the strong and the electromagnetic
amplitudes is large for $\jpsi\rightarrow1^-0^-$ decays.

 \item In the strong amplitude, the SU(3) breaking term $\epsilon$ 
is negative. 
\end{enumerate}

For $\psip$ decays, only two decay modes have been observed with 
finite branching ratios~\cite{bes}:
\begin{eqnarray*}
{\cal{B}}_{\KKSN+c.c.} &=& ( 0.81 \pm 0.24 \pm 0.16 ) \times 10^{-4}~~, \\
 {\cal{B}}_{\OP}  &=& ( 0.38 \pm 0.17 \pm 0.11 ) \times 10^{-4}~~.
\end{eqnarray*}
These are given without subtracting the contribution from $a_c$. 
In the following calculations, they are converted into measured cross
sections by multiplying the total $\psip$ cross section at $3.6861$ GeV for
the energy spread of 1.3 MeV~\cite{wymplb}. The $\OP$ mode gives the $\OP$
form factor at $\psip$
\beq
|{\cal F}_{\OP}(M^2_{\psip})|=(0.039^{+0.009}_{-0.012}) \mbox{GeV}^{-1}~~,
\label{Fpsip}
\eeq
which is related to $a_c$ and $\agcc$ by Eqs.~(\ref{ac}) and (\ref{ag}). 
Then it requires the input of $\phi$, together with the branching ratio
${\cal{B}}_{\KKSN+c.c.}$ to obtain 
$|a_{3g}+\epsilon|=|a_\gamma| {\cal R}{\cal C}$. 
Here four different phases are assumed: $+76.8^\circ$ and $-72.0^\circ$ from
$\jpsi$ fitting, $180^\circ$ from Ref.~\cite{suzuki}, and $-90^\circ$ which
is one of the phases favored by theory~\cite{gerard}. Then the cross
sections and the branching ratios of $\KKSC$ are calculated and listed in
Table~\ref{preditres}. Finally with the input of SU(3) breaking magnitude
${\cal R}=0.775\pm0.013$ from $\jpsi$ fitting, the cross sections of
$\rhopi$ are presented in Table~\ref{preditres}. For comparison, the upper
experimental limits of $\KKSC$ and $\rhopi$ cross sections are also listed.

\begin{table*}
\caption{\label{preditres} Calculated results for $\psip \ra \KKSC$ and
$\rho^0 \pi^0$ with different $\phi$.}
\begin{ruledtabular}
\begin{tabular}{cc||cc|cc}
$\phi$ {\rule[-3.5mm]{0mm}{7mm}}
         &${\cal C}= \left| {\displaystyle \frac{\aggg}{\agcc}} \right|$ 
             &$\sigma_{pre} (\KKSC) \mbox{(pb)} $
                 &${\cal B}^0_{\KKSC} (\times 10^{-5}) $\footnote{
\begin{minipage}{12.8cm}\mbox{}The supscript 0 indicates that
the continuum contribution in cross section has been subtracted.
\end{minipage}} 
                            & $\sigma_{pre} (\rho^0 \pi^0) \mbox{(pb)} $
			        &
${\cal B}^0_{\rho^0 \pi^0} (\times 10^{-5}) $        \\ \hline
$+76.8^{\circ}$ {\rule[-2mm]{0mm}{7mm}}
         &$7.0^{+3.1}_{-2.2}$
             &$37^{+24}_{-23}$
                 &$5.0^{+3.2}_{-3.1} $
	                    &$64^{+43}_{-41}$
                                &$9.0^{+6.1}_{-6.0} $       \\
$-72.0^{\circ}${\rule[-2mm]{0mm}{7mm}}
         &$5.3^{+3.1}_{-2.6}$
             &$19^{+14}_{-14}$
	         &$3.1^{+2.3}_{-2.3}$
			    &$33^{+25}_{-24}$
			        &$5.5^{+4.1}_{-4.0} $ \\
$-90^{\circ}$ {\rule[-2mm]{0mm}{7mm}} 
        &$4.5^{+3.1}_{-2.6}$
             &$12^{+9}_{-9}$
                 &$2.0^{+1.5}_{-1.5} $
                            &$22^{+17}_{-17} $
                                &$3.7^{+2.9}_{-2.9} $ \\
$180^{\circ}$ {\rule[-2mm]{0mm}{7mm}}
        &$3.4^{+3.0}_{-2.2}$
             &$4.0^{+4.3}_{-3.2}$
                 &$0.39^{+0.42}_{-0.31}$
                            &$7.8^{+8.6}_{-6.7}$
	                                   &$1.0^{+1.1}_{-0.8}$  \\  \hline
\multicolumn{2}{c||}{BES observed}
             &\multicolumn{1}{c}{$<9.6$}    
	         &\multicolumn{1}{c|}{}
			   &\multicolumn{1}{c}{$<5.8$}
			        &\multicolumn{1}{c}{}   \\
\end{tabular}
\end{ruledtabular}
\end{table*}

From Table~\ref{preditres}, it could be seen that one of the phases
$\phi=76.8^\circ$ gives poorer agreement than the other phase
$\phi=-72.0^\circ$ does. The former predicts cross sections for $\KKSC$ and
$\rhopi$ more than 1.2 and 1.4 standard deviations above the available upper
limits, while the latter yields 0.7 and 1.1 standard deviations from the
experimental upper limits.
This means, although theory and experiment at $\jpsi$ could not tell whether
the large phase is positive or negative, the measurements at $\psip$ favor the
negative one, which indicates destructive interference between the $\aggg$
and $\agee$ for $\rhopi$ and $\KKSC$, and constructive interference for
$\KKSN$ at the resonance. 

Compared with $\phi=-72.0^\circ$, the calculated cross sections with
$\phi=-90^\circ$ are closer to the measured upper limits for $\KKSC$ and
$\rhopi$, both of which are within one standard deviation. With
$\phi=180^\circ$, the evaluations on both cross sections also cover the
experimental upper limits within one standard deviation. 

From above analysis, it shows that the large negative phase
$-90^\circ$ suggested by theory and $-72.0^\circ$ from $\jpsi$ could fit
the $\psip$ data, after the one-photon annihilation amplitude being
considered properly. It should be noted that the $\rhopi$ cross sections in
Table~\ref{preditres} are calculated under the assumption that the SU(3)
breaking effect $\cal R$ has the same magnitude in $\jpsi$ and $\psip$
decays. If this assumption is removed, the $\rhopi$ cross section, for all
values of $\phi$, can be lower than those listed in Table~\ref{preditres}.
In the extreme situation in which $a_{3g}$ is very small, only $a_c$ and
$a_\gamma$ contribute, the $\rhopi$ cross section is 1/9 of $\OP$ cross
section, which is below the current upper limit by the experiment.

In conclusion, this work shows that the current available $\psip$ data 
accommodate both a large negative phase and $\phi=180^\circ$ with the
contribution from continuum one-photon annihilation amplitude $a_c$ taken
into account. The theoretical favored phase $-90^\circ$ could not be ruled
out as analyzed in Ref.~\cite{suzuki}.
The data are also consistent with the assumption that the SU(3) breaking
effect is of the same magnitude for $\jpsi$ and $\psip$ decays.
It requires more accurate $\psip\rightarrow1^-0^-$ data to determine 
the phase between $\aggg$ and $a_\gamma$. The most 
important information is whether the upper limits of $\KKSC$ and $\rhopi$
will be further pushed down, or finite cross sections will be observed. 

In the end, it is also interesting to notice that if
$\Gamma(\psip \rightarrow \gamma^* \rightarrow 1^-0^-)$/$\Gamma(\psip
\rightarrow ggg \rightarrow 1^-0^-)$ is roughly equal to 
$\Gamma(\psip\rightarrow\gamma^*\rightarrow X)/\Gamma(\psip\rightarrow ggg
\rightarrow X)$~\cite{suzuki}, then it is expected that 
${\cal C}\approx 4.5$, 
which is in good agreement with the fittings with $\phi=-72.0^\circ$,
$-90^\circ$ and $180^\circ$ in Table~\ref{preditres}; on the contrary,
similar relation for $\jpsi$ implies ${\cal C}\approx 4$, which is far less
than the fitted value ${\cal C}=10.3 \pm 0.3$. So far as this point is
concerned, the so-called ``$\rho\pi$ puzzle'' seems to be in $\jpsi$ decays
rather than in $\psip$. 

This work is supported in part by the National Natural Science Foundation of
China under contract No. 19991483 and 100 Talents Programme of the Chinese
Academy of Sciences under contract No. U-25.


\begin{thebibliography}{99}
\bibitem{refft1}In Ref.~\cite{suzuki}, the phase $\delta$ is defined
between the strong amplitude $a_1$ and $negative$ electromagnetic amplitude
$-a_\gamma$; while in this paper, the phase $\phi$ is defined between the
strong amplitude $\aggg$ and $positive$ electromagnetic amplitude $\agcc$. So 
$\delta=0^\circ$ in Ref.~\cite{suzuki} is equivalent to $\phi=180^\circ$  in
this paper.
\bibitem{refft2}Here $\aggg$ is $a_1$ defined in Ref.~\cite{suzuki} and the
definitions of $a_\gamma$ and $\epsilon$ are identical. In Ref.~\cite{bramon},
the notation $A$ and $A (\lambda-1)$ are equal to $a_{3g}$ and $2 \epsilon$ in 
this analysis, and $\lambda$ is associated to the mass spiltting between $s$ quark
and $u(d)$ quark by $\lambda = m_u/m_s=0.6$. In this scheme, the $\lambda$ 
(or $\epsilon$) has the same phase as $A$ (or $a_{3g}$).
\end{thebibliography}
\end{document}